\documentclass[dvips]{acta}
\usepackage{longtable}
\usepackage[small,bf]{caption}
\usepackage{rotating}
\usepackage{lscape,epsfig}
\usepackage{amssymb}
\usepackage{amsmath}
\DeclareSymbolFont{ppa}{OT1}{ppl}{m}{it}
\DeclareMathSymbol{\vv}{\mathalpha}{ppa}{'166}

\newfont{\hb}{rphvb at 10pt}
\newfont{\hbo}{rphvbo at 10pt}
\newfont{\bitt}{rptmbi at 12pt}
\newfont{\bits}{rptmbi at 11pt}

\SetPages{0}{0}

\SetVol{63}{2013}

\begin{document}
\newcommand{\TabApp}[2]{\begin{center}\parbox[t]{#1}{\centerline{
  {\bf Appendix}}
  \vskip2mm
  \centerline{\small {\spaceskip 2pt plus 1pt minus 1pt T a b l e}
  \refstepcounter{table}\thetable}
  \vskip2mm
  \centerline{\footnotesize #2}}
  \vskip3mm
\end{center}}

\newcommand{\TabCapp}[2]{\begin{center}\parbox[t]{#1}{\centerline{
  \small {\spaceskip 2pt plus 1pt minus 1pt T a b l e}
  \refstepcounter{table}\thetable}
  \vskip2mm
  \centerline{\footnotesize #2}}
  \vskip3mm
\end{center}}

\newcommand{\TTabCap}[3]{\begin{center}\parbox[t]{#1}{\centerline{
  \small {\spaceskip 2pt plus 1pt minus 1pt T a b l e}
  \refstepcounter{table}\thetable}
  \vskip2mm
  \centerline{\footnotesize #2}
  \centerline{\footnotesize #3}}
  \vskip1mm
\end{center}}

\newcommand{\MakeTableApp}[4]{\begin{table}[p]\TabApp{#2}{#3}
  \begin{center} \TableFont \begin{tabular}{#1} #4
  \end{tabular}\end{center}\end{table}}

\newcommand{\MakeTableSepp}[4]{\begin{table}[p]\TabCapp{#2}{#3}
  \begin{center} \TableFont \begin{tabular}{#1} #4
  \end{tabular}\end{center}\end{table}}

\newcommand{\MakeTableee}[4]{\begin{table}[htb]\TabCapp{#2}{#3}
  \begin{center} \TableFont \begin{tabular}{#1} #4
  \end{tabular}\end{center}\end{table}}

\newcommand{\MakeTablee}[5]{\begin{table}[htb]\TTabCap{#2}{#3}{#4}
  \begin{center} \TableFont \begin{tabular}{#1} #5
  \end{tabular}\end{center}\end{table}}

\def\thefootnote{\fnsymbol{footnote}}

\begin{Titlepage}
\Title{The Clusters AgeS Experiment (CASE): \\
 Variable Stars in the Globular Cluster 47 Tucanae
\footnote{Based on data obtained at Las Campanas Observatory using 6.5 m Magellan Clay
telescope and Swope 1.0 m telescope.}%
}
\Author{J.~~K~a~l~u~z~n~y$^1$,
      ~~M.~~R~o~z~y~c~z~k~a$^1$,
      ~~W.~~P~y~c~h$^1$,
      ~~W.~~K~r~z~e~m~i~n~s~k~i$^1$,
      ~~K.~~Z~l~o~c~z~e~w~s~k~i$^1$
      ~~W.~~N~a~r~l~o~c~h$^1$
      ~~and~~I.~B.~~T~h~o~m~p~s~o~n$^2$}
{ $^1$ Nicolaus Copernicus Astronomical Center, ul. Bartycka 18, 00-716
  Warsaw, Poland\\
  e-mail: (jka, mnr, wojtek, pych, kzlocz, wnarloch)@camk.edu.pl\\
  $^2$Carnegie Observatories, 813 Santa Barbara Street, Pasadena, CA 91101-1292, USA\\
     e-mail: ian@obs.carnegiescience.edu}
\Received{June 1, 2013}
\end{Titlepage}

\vspace*{7pt} \Abstract {Based on over 5400 $BV$ images of 47 Tuc
collected between 1998 and 2010 we obtained light curves of 65
variables, 19 of which are newly detected objects. New variables are
located mostly just outside of the core in a region poorly studied
by earlier surveys of the cluster. Among
them there are four detached eclipsing binaries and five likely
optical counterparts of X-ray sources. Two detached systems are promising
targets for follow-up observations. We briefly discuss properties of
the most interesting new variables.}
 {\it globular clusters: individual: 47 Tuc
 -- blue stragglers -- binaries: eclipsing -- stars: variables}
\section {Introduction}
47 Tuc (NGC~104), one of the brightest and most massive Milky Way's
globular clusters (GC), is located about 4.5~kpc away from the Sun at a high
galactic latitude of $-44.9$~deg (Harris 1996, 2010 edition). With a core
radius of 0$^{\prime}.$36 and a half-light radius of 3$^{\prime}.$17,
it had long been suspected to be on the verge of core-collapse; however
recent simulations of Giersz \& Heggie (2011) suggest that the collapse
may not take place for another $\sim$25 Gyr. In either case, the very
central region of 47 Tuc is unaccessible to the photometry with ground-based
telescopes lacking adaptive optics. The outer parts of the cluster, up to
a radius of about $\sim$22$^{\prime}$, can be observed rather easily, the
only obstacle being the contamination by stars from the outskirts of the
Small Magellanic Cloud.

Early photometric studies of 47 Tuc led to the detection of a number of
pulsating variables (Fox 1982 and references therein). An early spectroscopic
search for binary stars (Mayor et al. 1984) yielded a negative result: no
such objects were found among 64 stars located farther than 5 core radii from
the cluster center, "indicating a lower binary frequency and/or different
distribution of orbital parameters than in field stars in the solar
neighborhood". The first binaries in 47 Tuc (two W UMa type and six
detached or semidetached systems) were identified based on the HST-WFPC1
time-series photometry of cluster's core by Edmonds et al. (1996). Another
two W UMas were found by Kaluzny et al. (1997).
A more extensive survey of Kaluzny et al. (1998) resulted in the detection of
42 variables, among which there were 12 eclipsing binaries (nine W UMas and
three detached/semi-detached systems).

Using HST-WFPC2, Albrow et al. (2001) derived a time series
photometry for over 46,000 main-sequence stars in the core of 47
Tuc. Among them they found 11 detached eclipsing binaries, 15 W
UMas, 10 contact or near-contact non-eclipsing systems, and 71
variables with nearly sinusoidal to clearly non-sinusoidal light
curves and periods ranging from 0.4 to 10 d, which they classified
as BY Dra stars rotating synchronously with unseen companions. Six
of the latter were located far to the right of the cluster's main
sequence (MS), prompting the authors to define a new class of GC
members which they proposed to call red stragglers. An extensive
ground-based survey by Weldrake et al. (2004) led to the detection
of 69 new variables, more than doubling the number of such objects
in the 47 Tuc field. About 70\% of the newly identified variables
belonged to SMC.

One of the new detections of Weldrake et al. (2004) -- the detached
eclipsing binary V69 -- was subject to a detailed photometric and
spectroscopic analysis performed by Thompson et al. (2010). They
derived absolute parameters of its components with an accuracy
better than 1\%, estimated its age using mass-luminosity-age,
mass-radius-age, and (turnoff mass)-age relations, and determined
the helium abundance $Y$ with an accuracy of 0.03 for each of the
components. Having at least one detached eclipsing binary (DEB)
morewould enable a more accurate
direct determination of $Y$ for 47 Tuc. This exciting possibility
caused us to embark on an extensive survey aimed at the detection of
new DEBs in outer parts of 47 Tuc (note that objects located in the
outskirts of the cluster are only suitable for the ground-based
high-resolution optical spectroscopy).

Section 2 contains a brief report on the observations and explains
the methods used to calibrate the photometry. Variables (both the
newly detected and the known ones) identified during our survey are
described in Section 3. A summary of the paper is contained 
in Section 4.
\section {Observations and Photometric Reductions}
The cluster was observed at Las Campanas Observatory on the 1.0-m Swope telescope
equipped with the $2048\times 3150$ pixel SITe3 camera.
Two overlapping fields with the size $14^{\prime}.45$ by $22^{\prime}.84$ (henceforth
referred to as E and W) were covered. The longer axis of each field was oriented N-S.
The overlap region extended over $1^{\prime}.58$ in RA, so that the effectively
observed composite field had a size of $27^{\prime}.32$ by $22^{\prime}.84$. The
composite field was centered on the core of 47 Tuc.
All the images were obtained with the same set of $BV$ filters at a scale of
0.435~arcsec/pixel. Those taken at a seeing in excess of 3 arcsec were not
included in the present analysis.

The bulk of observations were collected during two long runs (43 nights in total)
in 2009 and 2010. We obtained then 2879 useful frames in $V$ and 344 in $B$ at a
median seeing of 1.5 and 1.6 arcsec, respectively, with average exposures of 117~s
for $V$ and 193~s for $B$. The exposure time of a given frame depended on the
momentary seeing, and was adjusted so as to keep the level of saturation constant
(i.e. to maintain the same magnitude of the brightest unsaturated stars).
The data collected in 2009 -- 2010 were supplemented with a set of images obtained
between 1998 and 2008 which comprised of 1337 $V$- and 400 $B$-frames of field E
together with 371 $V$- and 95 $B$-frames of field W. A part of that set was
subrastered to $14^{\prime}.45$ by $15^{\prime}.6$. The images had the same quality
as those collected in 2009 -- 2010.

The light curves of detected stars were extracted with the image subtraction
package DIAPL.\footnote{Freely accessible at
http://users.camk.edu.pl/pych/DIAPL/index.html}
Daophot, Allstar and Daogrow codes (Stetson 1987, 1990) were used to extract the
profile photometry of point sources and to derive aperture corrections for
reference images. The reference images in $V$ were made from 47 and 50
individual frames of fields E and W with an average seeing of 1.15 and 1.11
arcsec, respectively, while those in $B$ were made from 15 and 22 images, respectively,
with an average seeing 1.29 in both E and W.

For the analysis, the E and W fields were divided into $4\times 6$ segments to
reduce the effects of PSF variability. The light curves derived with DIAPL were
converted from differential counts to magnitudes based on profile photometry and
aperture corrections determined for each segment of reference images separately.
Instrumental magnitudes were transformed to the standard $BV$ system as described
in Section 2.1. The $V$-band light curves were extracted for 63024 sources in
field E and 52591 sources in field W. Accounting for the overlap, we obtained
110141 unique light curves in the whole composite field.

In Fig.~1 the $rms$ values of individual measurements in $V$ are plotted versus
the average $V$-magnitude for each object in field W. The
photometric accuracy reaches about 3 mmag at $V=14.0$ mag,
decreasing to $\sim$10 mmag for the turnoff stars at $V=17.0$ mag
and to $\sim$100 mmag at the faintest stars for which the
measurement could still be performed ($V=20.5$~mag). Exactly the
same accuracy was reached in field E. Stars with $V<13.0$ mag were
overexposed on the reference images of both fields, making the study
of their light curves impossible. Depending on exposure time and
seeing, some images of stars with 13$<V<13.5$ mag were also
saturated. Their light curves were filtered to remove points
affected by saturation.

\subsection{Calibration}
The instrumental magnitudes for reference images were converted to
the  standard ones using linear transformations based on 817 and 570
local Stetson's standards in fields W and E, respectively (Stetson
2000; 2009 on-line version). The following transformations were
derived:
\begin{eqnarray}
{\rm v=V -0.086(3)\times (B-V) -1.591(3)}\nonumber \\
{\rm b=B -0.146(4)\times (B-V) -2.054(4)} \\
{\rm b-v=0.938(3)\times (B-V) -0.462(3)}\nonumber
\end{eqnarray}
(field E) and
\begin{eqnarray}
{\rm v=V -0.074(3)\times (B-V) -1.602(3)}\nonumber \\
{\rm b=B -0.139(4)\times (B-V) -2.038(4)} \\
{\rm b-v=0.936(3)\times (B-V) -0.437(3)~}\nonumber
\end{eqnarray}
(field W), where capital letters denote the standard magnitudes.

As most of the monitoring was conducted in $V$ only, the standard magnitudes
were derived using average values of $B-V$. This introduced some systematic
errors in the case of color-changing objects. However, for most of the
variables the amplitude of color variations did nor exceed 0.1 mag,
and the systematic error of $V$ magnitude was smaller than 0.01 mag.
For the main goal of our survey, which is is to detect
new variables, inaccuracies of this size are unimportant.

The astrometric solutions for the reference images in $V$ were found
based on positions of 2997 UCAC3 stars (Zacharias et al. 2010). The
average residuals in RA and DEC between cataloged and recovered
coordinates amount to 0.14 and 0.14 arcsec, respectively. For the
detection of variables we used methods described in Kaluzny et al.
(2013). A total of 65 variables were found (among them 19 new ones).
They are listed in Table 1 along with their equatorial coordinates
and cross-identifications for catalogs published by Weldrake et al.
(2004), Kaluzny et al. (1998) and Samus et al. (2009). We detected
variability of all but one object from Weldrake et al. (2004)
present in the surveyed field (our photometry shows no brightness
variations for their star V93. 

Finding charts for newly detected variables are shown in Fig. 2.

\section{Variables}

The basic properties of variables detected in our survey are listed
in Table 2. The periods in column 2 were derived with the method
employing periodic orthogonal polynomials to fit the observations
and the analysis of variance statistic to evaluate the quality of
the fit (Schwarzenberg-Czerny 1996), as implemented in the TATRY
code kindly made available by the author.

The parameter $\Delta V$ listed in column 4 gives the full range of
the measured $V$-magnitude, including seasonal changes of light curves. The last
column of Table 2 contains the proposed classification of variables,
with ``new'' indicating newly detected objects. EW, EB and EA denote
eclipsing binaries with light curves of W~UMa, $\beta$ Lyrae and
Algol type, respectively; Ell stands for ellipsoidal variables, and
stars located along or near the red giant branch on the cluster
color-magnitude diagram (CMD) are labeled with RGB / RS CVn. The variability
of the latter is likely caused by ellipsoidal effect and/or
chromospheric activity related to binarity. Among the new variables
there are five likely optical counterparts to X-ray sources detected
with the Chandra telescope by Heinke et al. (2005).

Figure 3 presents the CMD of a small section of the monitored field
with marked locations of all but two variables listed in Tables 1
and 2. Not included are W26 and E20 - very red Miras with $B-V>3$
mag which were originally detected by the OGLE group (Soszynski et al.
2011). The group of variables located to the blue of the cluster's
main sequence at $V\approx19$ is composed of RR~Lyr pulsators from the
SMC. The sequence of red long period variables starting at
($V\approx19$, $B-V\approx1.1$) and extending to $B-V\approx2.0$
also consists of SMC stars. Fig. 4 shows the positions in the 
cluster CMD  of a  selection of
our variables, including all eclipsing binaries and three
particularly interesting objects which will be discussed below.
Phased light curves for 14 new variables are
displayed in Fig.~5 (not shown are RR-Lyrae stars, long period
variables, and E32 whose period is not known). Light curves for all
65 variables detected within the present survey can be found in the
electronic version of this paper available from the Acta Astronomica 
archive or from CASE archive at http://case.camk.edu.pl. 

\subsection{Eclipsing binaries}
The main result of our survey is the detection of four new detached
eclipsing binaries. All these objects are located beyond the core region of
the cluster and are suitable for spectroscopic follow-up studies
with ground-based telescopes.

The light curve of W12 shows a shallow secondary eclipse with $\Delta V
\approx 0.1$~mag and a primary eclipse with $\Delta V \approx 0.4$~mag.
A preliminary solution of the $V$-light curve indicates a high luminosity
ratio of the components: $L_{p}/L_{s}=9$. The expected luminosity ratio
for the $I$ band equals to 6, so that the near-IR spectroscopy may allow
for the determination of radial velocity curves of both components.

The available light curve of E32 shows only one clear eclipse
with $\Delta V=0.6$ mag. This means that the orbital period has to
be longer than 10 days (and may be significantly longer). We
obtained a single spectrum of E32 with the MIKE Echelle spectrograph
on  the 6.5 m Magellan Clay telescope, and we found that the
binary is an SB2 system. Several additional spectra are needed to
establish the ephemeris with confidence. This in turn will enable
further photometric observations in eclipses, and finally the
determination of absolute parameters of the system. The location of
E32 on the CMD indicates that at least one of its components is an
evolved star at or past the turnoff. The analysis of E32 together
with V69 (Thompson et al. 2010) may allow interesting limits 
to be placed on the helium abundance of 47~Tuc.
\begin{table}
\caption{Equatorial coordinates of variable stars identified
within the present survey \label{tab:coords}}
{\scriptsize
\setlength{\tabcolsep}{0.48em}
\begin{tabular}{|l|c|c|c|c|c|l|c|c|c|c|c|}
\hline
ID &  RA$_{J2000}$& Dec$_{J2000}$ & ID-W$^{a}$ & ID-K$^{b}$ & ID-S$^{c}$& ID &
      RA$_{J2000}$& Dec$_{J2000}$ & ID-W$^{a}$ & ID-K$^{b}$ & ID-S$^{c}$\\
      &  [deg]    & [deg]   &            &            &           &          & [deg]      &     [deg]  &            &         \\
\hline
\hline
W1  &6.04262 &-72.06708& & &           &E8  &6.51136 &-71.97276& &231  & \\
W2  &5.91532 &-72.02788& & &           &E9  &6.45689&-71.97410& V99 &226&        \\ 
W3  &6.05276 &-72.11105& & &V001       &E10 &6.55163&-72.03701& V27 &230&        \\  
W4  &6.02083 &-72.08958& & &           &E11 &6.51274&-72.05086& V28 &229&        \\
W5  &5.99704 &-72.11350& & &           &E12 &6.38434&-72.03098& V100&225  & V044 \\
W6  &5.91897 &-72.09997& &217 &V009    &E13 &6.53692&-72.11706& V7  &228  & V046 \\
W7  &5.95557 &-72.21893& & &           &E14 &6.42508&-72.10040& V10 &223  &      \\
W8  &5.69983 &-71.94696& & &           &E15 &6.54387&-72.18549& V6  &227  & V045\\
W9  &5.68282 &-71.95575& V71&234 &     &E16 &6.53915&-72.20798& V5  &255  & \\
W10 &5.77913 &-72.02260&  &218&        &E17 &6.39210&-72.16616&    V9  &222  & \\
W11 &5.84984 &-72.05125& & &           &E18 &6.51177&-72.24155&     &     & \\
W12 &5.73147 &-72.08756& & &           &E19 &6.40987&-72.23572& & & \\
W13 &5.72354 &-72.06296& V69& &        &E20 &6.40448&-72.25761& & 252   & \\
W14 &5.65323 &-72.11997& & &           &E21 &6.31753&-71.93490& V30 & 238 &V047\\
W15 &5.81761 &-72.18639& & &           &E22 &6.24824&-71.89658& & 241   & \\
W16 &5.77493 &-72.15847& V97& &        &E23 &6.20544&-71.93885& V34 &   & \\
W17 &5.67023 &-72.15625& V96  &215&    &E24 &6.25194&-72.00076& V32 &221&V043\\
W18 &5.69786 &-72.22143& V95&245&V050  &E25 &6.17727&-72.10601& & & \\
W19 &5.63677 &-71.99193& V70&216&      &E26 &6.17207&-72.10331& & & \\
W20 &5.54779 &-71.98831& V62 & &       &E27 &6.29647&-72.20400& V14&250 &V052\\
W21 &5.50268 &-72.03445& V61 &214&V042 &E28 &6.27575&-72.17319& V15&219 & \\
W22 &5.46433 &-72.18055& V91 & &       &E29 &6.15076&-71.95535&  V33&239 \\
W23 &5.63907 &-72.23045& V94 & 246&    &E30 &6.17717&-71.98994& V31&220 & \\
W24 &5.27715 &-71.94372& V64& &        &E31 &6.05647&-71.96737& & & \\
W25 &5.34974 &-72.25939& V92&243&      &E32 &6.15633&-72.06476& & & \\
W26 &5.27031 &-72.25646&    &242&      &E33 &6.07805&-72.08192& & & \\
E1  &6.75007 &-71.91918& V26 & &       &E34 &6.10767&-72.11765& & & \\
E2  &6.69230 &-71.99538& & &           &E35 &6.09773&-72.14155& V17 & & \\
E3  &6.73688 &-72.17036& V13 & 232 &   &E36 &6.09675&-72.12296& & & \\
E4  &6.69681 &-72.27622& & &           &E37 &6.07745&-72.13304& & &V002\\
E5  &6.67914 &-72.25578& V12 & 253&V053 &E38 &6.13241&-72.15819& V16&251 & \\
E6  &6.58064 &-72.23394& V4& 254 &     &E39 &6.10552&-72.22732& & & \\
E7  &6.40631 &-71.93440& V29 & &       &    &       &           &  &\\
\hline
\end{tabular}
}
{\footnotesize$^a$ Weldrake et al. 2004;
$^b$ Kaluzny et al. 1988; $^c$ Samus et al. 2009}
\end{table}

For W7 we have three MIKE/Magellan spectra. They show that the
variable is an SB1 binary with  measured velocities of
$-37.41\pm0.32$, $-74.27\pm0.22$ and $+16.38\pm0.16$ km/s at orbital
phases 0.49 (HJD 245 5769.866), 0.353 (HJD 245 5770.839) and 0.692 
(HJD 55836.642), respectively. 
These measurements indicate that the binary is a
likely member of the cluster. The systemic velocity of 47 Tuc is
equal to -18.0 km/s, and the velocity dispersion at the location 
of W7 amounts to $\sim$11 km/s (Harris 1996). This can be
compared with the binary's velocity near conjunction at orbital phase 0.49.
The fourth detached system, E39, is placed rather far
from the cluster main sequence and may be a foreground halo object.
Its low luminosity and short orbital period make it a
difficult target for spectroscopy.

\begin{center}
\footnotesize{
\begin{longtable}{|l|l|c|c|c|c|}
\caption[]{Properties of variable stars identified
within the present survey}\label{tab:properties}\\
\hline
ID &  P       & $V_{max}$ & $\Delta V$ & $<B-V>$ & Type of variability \\
      &  [d]  & [mag]     & [mag]      &         &    Remarks   \\
\hline \hline
\endfirsthead
\multicolumn{6}{l}
{{ \tablename\ \thetable{} -- continued from previous page}} \\
\hline
ID &  P       & $V_{max}$ & $\Delta V$ & $<B-V>$ & Type of variability \\
      &  [d]  & [mag]     & [mag]      &  [mag]  &  Remarks     \\
\hline
\hline
\endhead
\hline \multicolumn{6}{|r|}{Continued on next page} \\ \hline
\endfoot

\hline
\endlastfoot
W1 &   0.36115544(1) & 15.607&  0.163 & 0.600 & EW, YS, X$^{a}$, {\bf new}, {\scriptsize Ch-0024410.2-720401}\\
W2 &   0.28287568(1) & 17.565 & 0.317 & 0.543 & EW, {\bf new} \\
W3 &   -             & 11.753 & 2.872 & 1.516 & LP \\
W4 &   0.35298992(1) & 15.838 & 0.163 & 0.271 & EW, BS, X, {\bf new}, {\scriptsize Ch-002404.9-720522}\\
W5 &   0.28036018(8) & 17.143 & 0.225 & 0.420 & EW, BS, X, {\bf new}, {\scriptsize Ch-002359.3-720648}\\
W6 &   0.73703372(1) & 13.063 & 1.024 & 0.261 & RR, {\scriptsize OGLE-SMC-RRLYR-0051}$^{~b}$\\
W7 &   3.88255(2)    & 19.091 & 0.384 & 0.815 & EA, {\bf new} \\
W8 &   12.78227(2)   & 14.150 & 0.032 & 1.173 & RS CVn, RGB, {\bf new} \\
W9 &   0.6158898(1)  & 19.084 & 0.811 & 0.439 & RR, {\scriptsize OGLE-SMC-RRLYR-0030} \\
W10&   -             & 15.976 & 0.384 & 1.652 & LP, {\scriptsize OGLE-SMC-LPV-00128}$^{~c}$ \\
W11&   0.63228616(9) & 18.948 & 0.65  & 0.18  & RR, {\scriptsize OGLE-SMC-RRLYR-0044}  \\
W12&   3.732001(2)   & 17.462 & 0.411 & 0.634 & EA, {\bf new} \\
W13&   29.53975(1)   & 16.799 & 0.616 & 0.537 & EA \\
W14&   0.04595852(3) & 14.679 & 0.037 & 0.145 & SX, BS, {\bf new}$^d$ \\
W15&   71.4156(5)    & 17.138 & 0.141 & 1.557 & LP, {\scriptsize OGLE-SMC-LPV-00143} \\
W16&   0.39708061(4) & 18.859 & 0.274 & 0.834 & EB \\
W17&   8.4278094(1)  & 16.638 & 0.203 & 0.931 & BY~Dra \\
W18&   0.27890017(1) & 15.459 & 0.424 & 0.565 & EW, YS             \\
W19&   0.36164588(6) & 19.584 & 0.624 & 0.276 & RR, {\scriptsize OGLE-SMC-RRLYR-0026} \\
W20&   66.6          & 16.991 & 0.611 & 1.792 & LP, {\scriptsize OGLE-SMC-LPV-00066} \\
W21&   0.273741797(2)& 17.919 & 0.360 & 0.610 & EW \\
W22&   0.65390427(7) & 18.738 & 0.264 & 0.582 & RR (blend?), {\scriptsize OGLE-SMC-RRLYR-0016} \\
W23&   0.57218584(4) & 19.017 & 0.980 & 0.428 & RR, {\scriptsize OGLE-SMC-RRLYR-0027} \\
W24&   0.595771(2)   & 19.323 & 0.839 & 0.301 & RR, {\scriptsize OGLE-SMC-RRLYR-0011} \\
W25&   0.6256305(15) & 19.364 & 0.752 & 0.404 & RR, {\scriptsize OGLE-SMC-RRLYR-0012} \\
W26&   269.3         & 16.879 & 1.406 & 3.347 & LP, {\scriptsize OGLE-SMC-LPV-00015} \\
E1 &   0.347489131(4)& 17.025 & 0.485 & 0.493 & EW, BS  \\
E2 &   0.8102273(2)  & 18.306 & 0.168 & 0.734 & RR, {\bf new} \\
E3 &   0.36349524(3) & 19.129 & 0.733 & 0.240 & RR, {\scriptsize OGLE-SMC-RRLYR-0122} \\
E4 &   79.92         & 16.850 & 0.297 & 2.053 & LP, {\scriptsize OGLE-SMC-LPV-00446} \\
E5 &   0.44625922(1) & 16.659 & 0.398 & 0.434 & EB, BS \\
E6 &   -             & 16.539 & 0.254 & 1.783 & LP, {\scriptsize OGLE-SMC-LPV-00398} \\
E7 &   4.463003(6)   & 18.792 & 0.210 & 1.273 & ? \\
E8 &   6.371513(8)   & 14.141 & 0.152 & 0.760 & RS CVn, RGB-clump          \\
E9 &   0.64752185(2) & 18.959 & 0.852 & 0.446 & RR, {\scriptsize OGLE-SMC-RRLYR-0098} \\
E10&   4.78667(2)    & 17.474 & 0.603 & 0.917 & BY~Dra \\
E11&   8.3711839(8)  & 14.916 & 0.157 & 0.911 & RS CVn, RGB \\
E12&   0.234635701(3)& 19.441 & 0.787 & 0.981 & EW \\
E13&   1.150686034(3)& 15.797 & 0.417 & 0.233 & EB, BS \\
E14&   0.29712114(1) & 17.306 & 0.505 & 0.238 & RR, {\scriptsize OGLE-SMC-RRLYR-0098} \\
E15&   0.378854028(1)& 16.397 & 0.311 & 0.349 & EW, BS \\
E16&   0.52515175(1) & 18.977 & 1.254 & 0.422 & RR, {\scriptsize OGLE-SMC-RRLYR-0104} \\
E17&   20.791(1)     & 16.576 & 0.207 & 0.733 & RS CVn, RGB  \\
E18&   14.09425(2)   & 17.176 & 0.070 & 1.405 & ?, {\bf new} \\
E19&   -             & 14.344 & 0.194 & 0.989 & LP, {\bf new} \\
E20&   266.4         & 17.911 & 2.254 & 3.607 & LP, {\scriptsize OGLE-SMC-LPV-00329} \\
E21&   0.250546155(3)& 18.397 & 0.457 & 0.669 & EW \\
E22&   35.3519(1)    & 16.747 & 0.111 & 1.680 & LP, {\scriptsize OGLE-SMC-LPV-00271} \\
E23&   0.24169575(1) & 18.551 & 0.337 & 0.693 & EW \\
E24&   0.313434596(1)& 17.650 & 0.519 & 0.511 & EW \\
E25&   3.33587(4)    & 17.849 & 0.143 & 0.589 & ?, {\bf new} \\
E26&   1.4043996(8)  & 18.528 & 0.282 & 1.276 & ?, {\bf new} \\
E27&   0.351384786(1)& 16.244 & 0.246 & 0.318 & EW, BS \\
E28&   36.8914(12)   & 15.252 & 0.148 & 0.900 & RS CVn, RGB \\
E29&   40.39(16)     & 16.613 & 0.111 & 1.608 & LP, {\scriptsize OGLE-SMC-LPV-00235} \\
E30&   10.77526(2)   & 16.007 & 0.320 & 0.814 & RS CVn, RGB \\
E31&   20.43463(6)   & 16.564 & 0.457 & 0.901 & RS CVn, RGB, {\bf new} \\
E32&   -             & 17.109 & 0.362 & 0.569 & EA, {\bf new}  \\
E33&   0.25151879(1) & 16.391 & 0.107 & 1.044 & Ell?, X, {\bf new}, {\scriptsize Ch-002418.6-720455}\\
E34&   9.87379(4)    & 15.495 & 0.294 & 0.878 & RGB, X, {\bf new}, {\scriptsize Ch-002425.8-720703 }\\
E35&   0.300361921(7)& 18.021 & 0.345 & 0.648 & EW \\
E36&   0.279869331(2)& 18.053 & 0.641 & 0.703 & EW, {\bf new} \\
E37&   135.43559(2)  & 13.180 & 2.480 & 1.301 & LP \\
E38&   3.480885(3)   & 16.468 & 0.313 & 0.843 & RS CVn, RGB \\
E39&   0.987519(2)   & 18.792 & 0.507 & 0.937 & EA, {\bf new} \\
\end{longtable}
}
\end{center}
\vskip -1cm
{\footnotesize
$^a$ X - likely counterpart of an X-ray source 
Ch-nnnnnnn.n-nnnnnn cataloged by Heinke et al. (2005)\\
$^b$ OGLE-SMC-RRLYR-nnnn = stars cataloged by Soszynski et al. (2010)\\
$^c$ OGLE-SMC-LPV-nnnnn = stars catalogud by Soszynski et al. (2011)\\
$^d$ independently discovered by Poleski (2012)}

In addition to the five detached binaries our sample includes one
semi-detached system (E13; Kaluzny et al. 2007) and 15 contact or
nearly-contact binaries. Six of them are blue stragglers, and two
(W1 and W18) are located in the region occupied by yellow stragglers
(see e.g. Stetson 1994). Variable yellow stragglers are rare objects
and therefore we examined HST/ACS images of W1 and W18 to check 
whether or not 
the ground based photometry is affected by blending. We found
W18 to be an isolated object with no indication for unresolved
visual companions on ACS images. Still, we cannot rule out the
possibility that it is a triple system like many (perhaps most) W
UMa stars (Rucinski et al 2007). Spectroscopic observations can
clarify this issue. As for W1, ACS images show three visual
components forming a blend which cannot be resolved on our ground
based images. Two closest components of the blend are separated by
0.3 arcsec. Thus, based on the available evidence, W1 cannot be
regarded as a yellow straggler.

\subsection{Notes on individual objects}
Some of the newly detected variables deserve a short comment on
their properties. The red straggler candidate E33 is located only 61 arcsec
away from the cluster centre; however it is an isolated object whose
ground based photometry is free from blending-related problems. This
was confirmed by the examination of HST/ACS/F625W images. The star
has a sine-like light curve with  $\Delta V=0.11$ mag and a period
of 0.25~d which was coherent and stable during several observing
seasons between 1998 and 2010. This indicates that the observed
variability is very likely related to the binarity of E33. We have
tentatively classified E33 as an ellipsoidal variable. Heinke et al.
(2005) analyzed two sets of Chandra data for 47 Tuc. The X-ray
counterpart of E33 was detected in 2002 with an X-ray luminosity of
$0.7$E30 erg/s in 0.5-2.5~keV band, but no X-rays were detected at
E33 position in 2000. This implies a seasonal variability of the
X-ray source connected with the star. The variable is too red to be
an ordinary contact binary regardless of its membership status in 47
Tuc. This is evident by comparing its color with colors of contact
systems E21 and E23 whose orbital periods are similar to the period
of E33 (see Fig. 4). As for the membership status of E33, the
examination of stacked subtracted images from seasons 1998, 1999,
2009, and 2010 does not indicate any proper motion with respect to
cluster stars. The qualitative method we used to estimate the proper
motion is based on Eyer and Wozniak (2001). Spectroscopic
data are needed to clarify the evolutionary and membership status of
E33. 

The variable E8 is located on the red horizontal branch on
the cluster CMD. The light curve shows coherent changes with $\Delta
V= 0.15$~mag and $P=6.4$~d. The amplitude of the light curve
exhibits small but easily visible seasonal changes. Similar
variations with the same period were detected by the OGLE group in
1993 (Kaluzny et al. 1998). Given the coherence of the variability
it is likely that the star is a binary. If so, it would be a rare
example of a photometrically variable binary from the red horizontal
branch. With $P=6.4$~d and a radius of the red giant component of
about 10~$R_{\odot}$, E8 would have to be a rather compact system.
Obviously, if the variability was induced mainly by the ellipsoidal
effect, the actual orbital period of a binary would double to
12.8~d.
The bright blue straggler W14 turned out to be an SX~Phe-type
pulsator. It is one of a few such variables detected in 47 Tuc
(Gilliland et al. 1998; Poleski 2012). 
SX Phe stars are
common among blue stragglers in metal-poor globular clusters, but
rather rare in metal rich ones.

Finally, we note the presence of several variables located on or
slightly to the red of the red giant branch of the cluster, newly 
detected of which are E31 and E34 (see Figs. 3 and 4). They are 
good candidates for binaries and are easy targets for a spectroscopic 
follow-up.
\section{Summary}
We performed a photometric study of the globular
cluster 47 Tuc with a time baseline of more than ten years. 
Based on over 6500 $BV$ frames we have identified 65
variable stars, 19 of which are new detections. We provide celestial
coordinates of all variables, and cross-identifications of the
variables discovered earlier by other authors. Finding charts for
the new variables are also provided. Five of the new variables are
likely optical counterparts of X-ray sources, and another four ones
are detached eclipsing binaries. Two detached eclipsing systems are
located close to the main-sequence turnoff on the CMD of 47 Tuc, and
we argue that they are promising targets for detailed photometric
and spectroscopic studies: when combined with the results of an
earlier study of W13 (Thompson et al. 2010; their variable V69) they
may allow an improved constraint on the helium content of the cluster.
The yellow straggler W18, the red straggler E33 and the red
horizontal branch object E8 are another systems deserving further
study which would clarify their membership and evolutionary status.

\Acknow{We are grateful to Igor Soszynski for a very detailed and helpful
referee report, and to Radek Poleski for pointing out the references to 
SX Phe stars in 47 Tuc. JK, MR, WP and WN were partly 
supported by the grant NCN 2012/05/B/ST9/03931 from the Polish Ministry of Science.}


\vskip 5cm

\begin{figure}[htb]
\centerline{\includegraphics[width=120mm, bb= 67 270 494 497,clip]{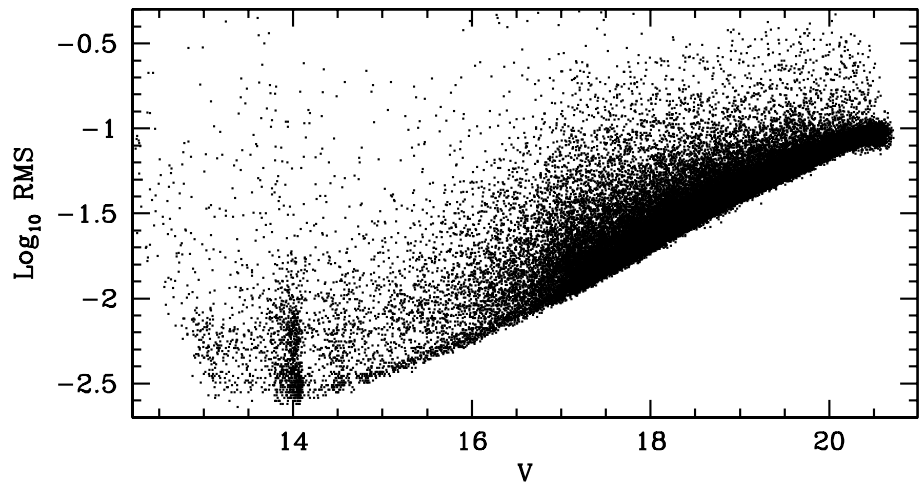}}
\caption{\small
The accuracy of the photometry of 47 Tuc. For each object in field W the {\it RMS}
values of individual measurements in the $V$-band are plotted vs. the average $V$-magnitude.}
\end{figure}

\clearpage

\begin{figure}[htb]
\centerline{\includegraphics[width=\textwidth, bb= 13 13 885 506, clip]{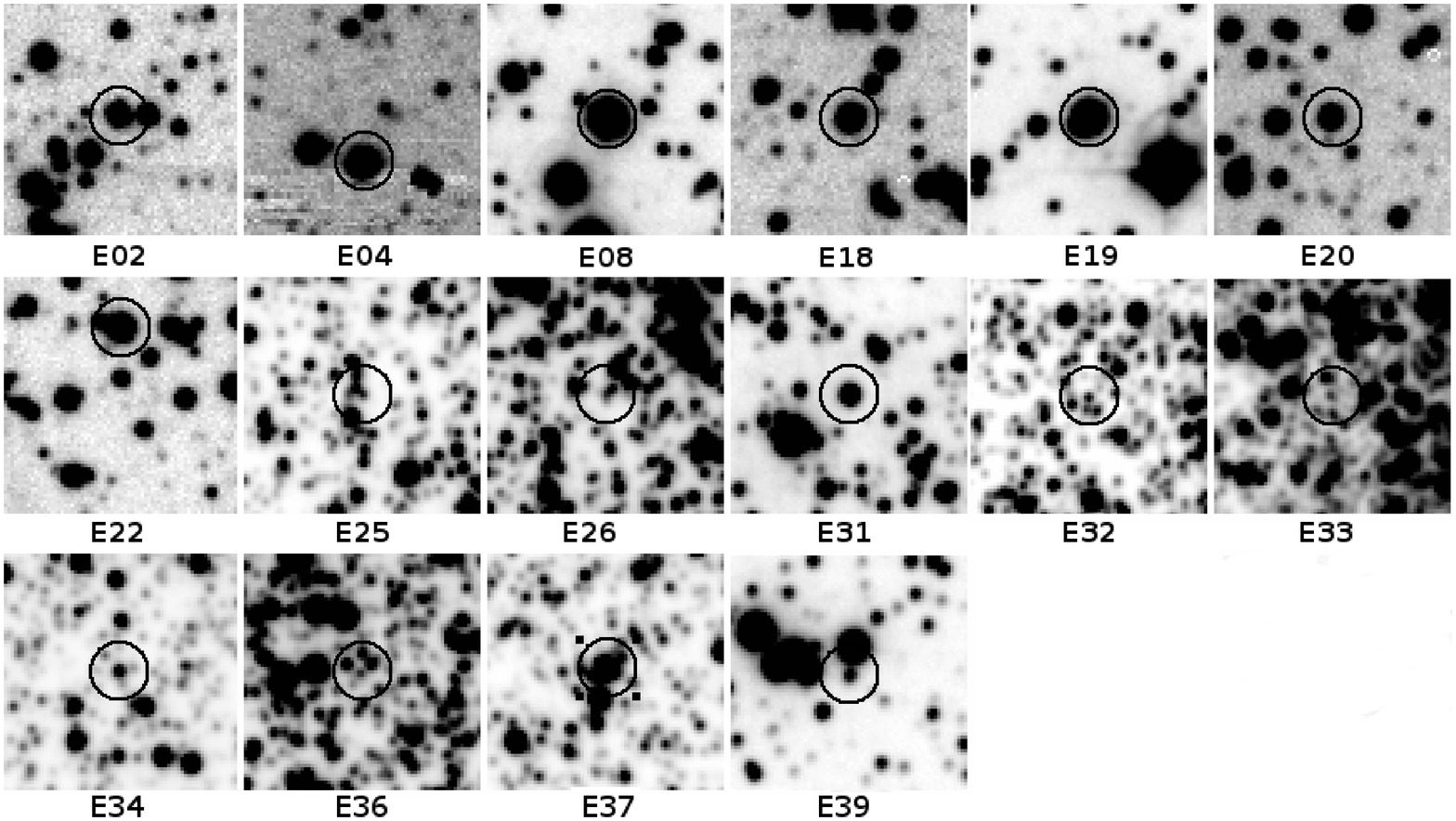}}
\end{figure}
\vspace{-1em}
\begin{figure}[h]
\centerline{\includegraphics[width=\textwidth, bb= 13 13 895 515 clip]{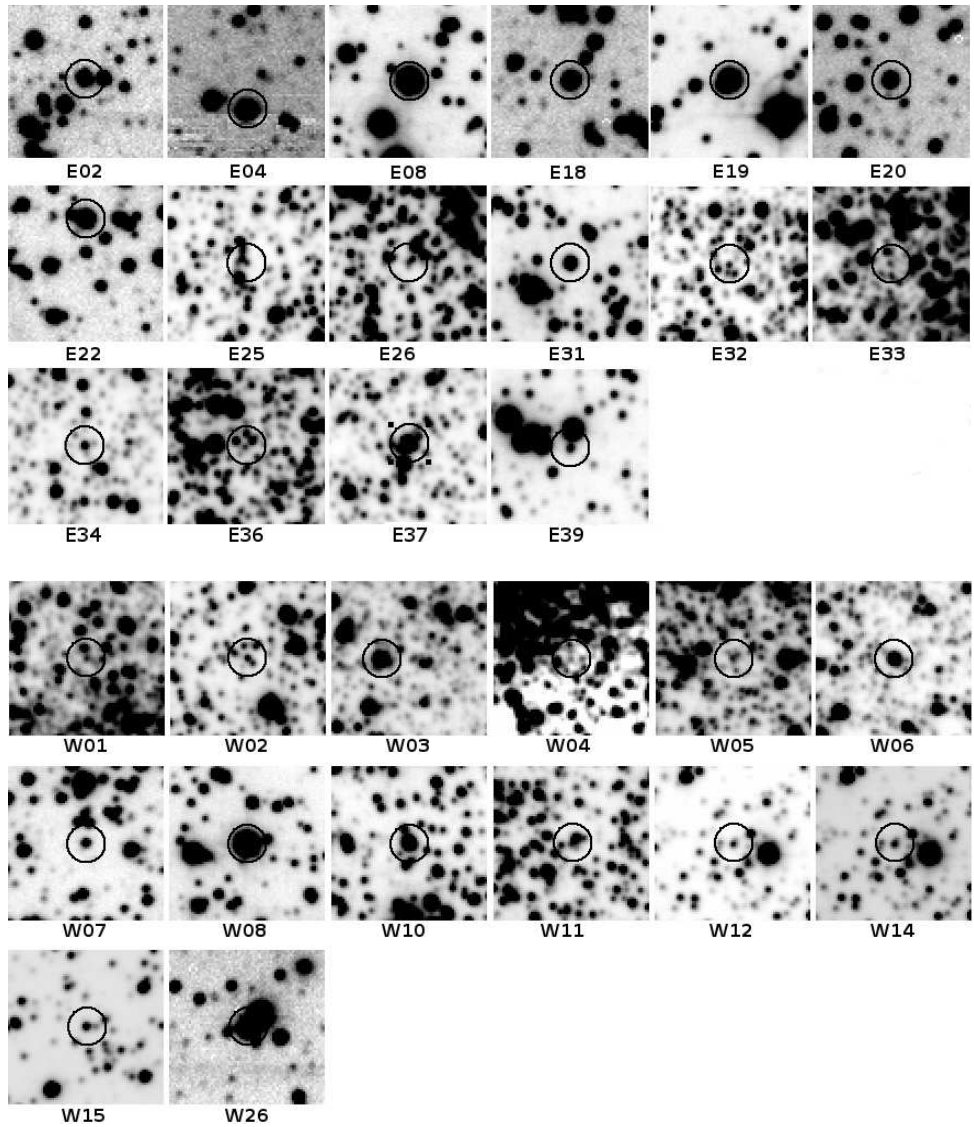}}
\caption{\small
Finding charts for the newly detected variables. Each chart is 30
arcsec on a side, with north up and east to the left.}
\end{figure}

\clearpage

\begin{figure}[htb]
\centerline{\includegraphics[width=120mm, bb= 54 169 564 689,
clip]{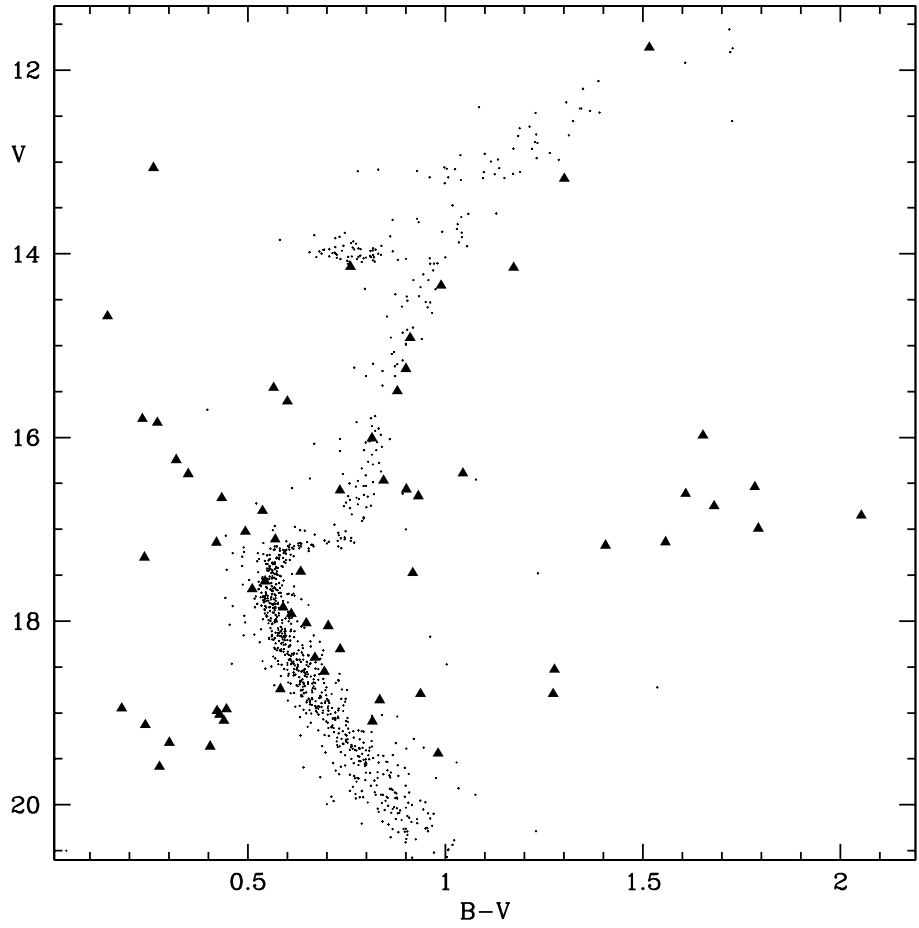}} \caption{\small CMD of 47~Tuc with positions of 63
variables detected by the CASE program marked with triangles.
Not shown are two very red Miras with $B-V>3$, 
detected earlier by the OGLE group (Soszynski et al. 2011).
Background dots: nonvariable stars selected from a small fragment of
the observed field.} 
\end{figure}

\clearpage

\begin{figure}[htb]
\centerline{\includegraphics[width=120mm, bb= 55 169 564 690,
clip]{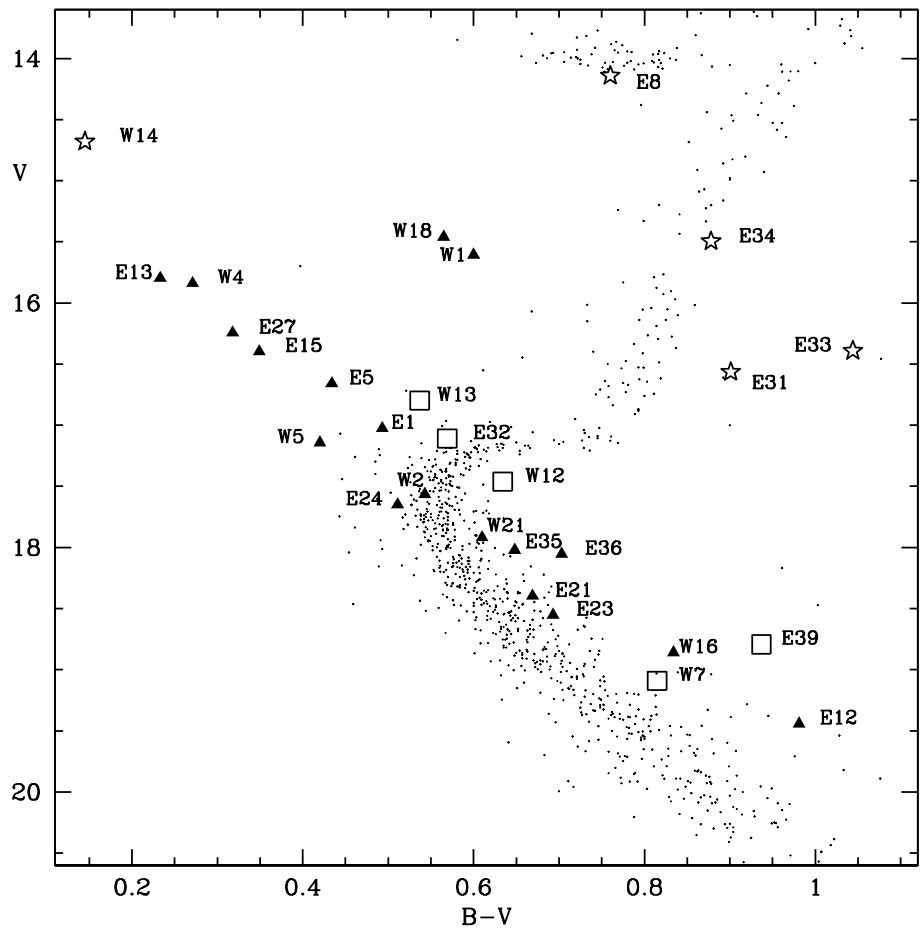}} \caption{\small CMD of 47~Tuc with positions of 
all binaries detected by the CASE program. Also shown  
are a few other variables briefly discussed in Section 3.2. Squares: 
detached binaries; triangles: contact binaries; stars: other variables.
Background dots: nonvariable stars selected from a small fragment of
the observed field.}
\end{figure}

\begin{figure}[htb]
\centerline{\includegraphics[width=\textwidth, bb= 58 152 517 707,
clip]{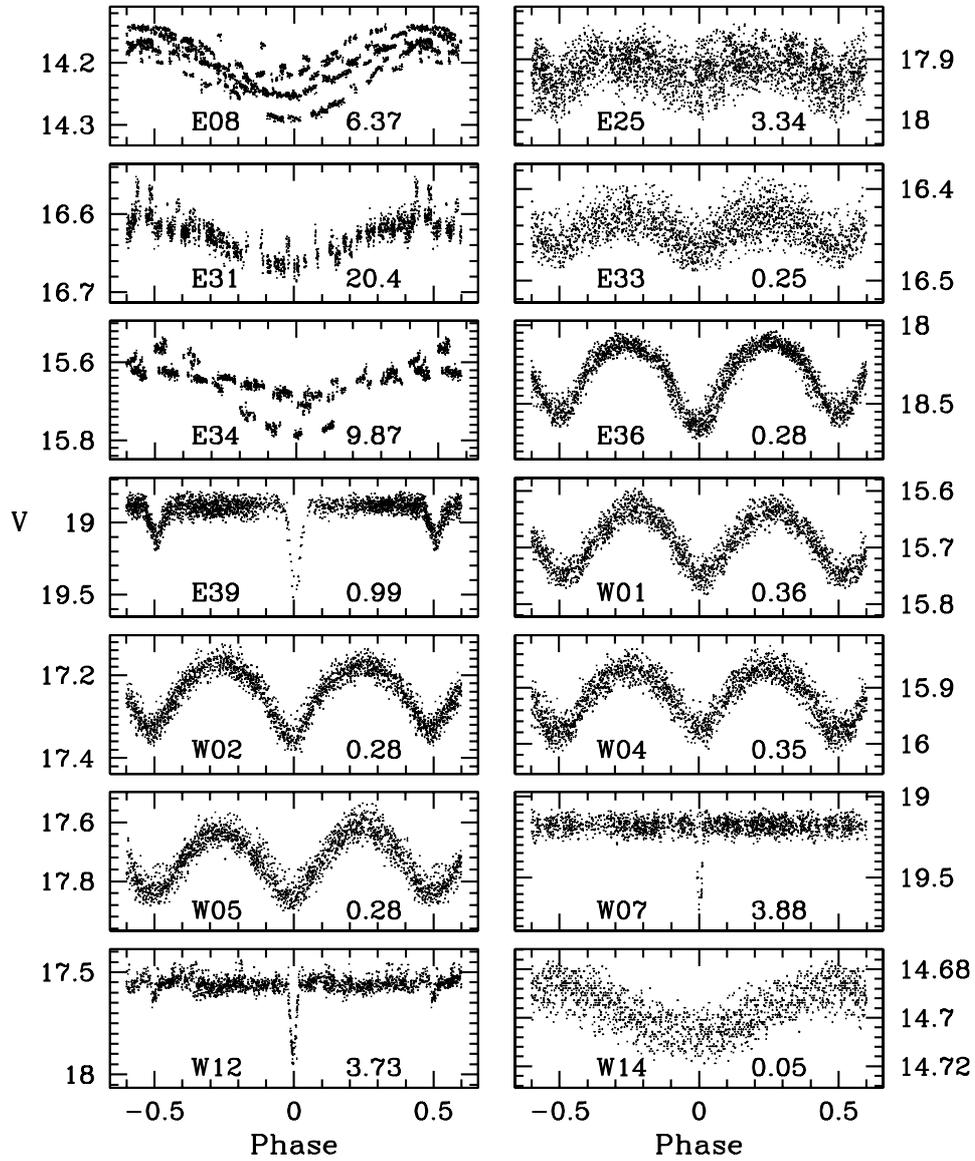}} \caption{\small Phased $V$-light curves of newly
detected eclipsing binaries and a few new variables of other types.
In each panel the left label gives the name of the variable, and the
right label the orbital period in days. }
\end{figure}

\end{document}